\documentclass[prd, superscriptaddress, showpacs, floatfix, notitlepage]{revtex4-1}
\usepackage{hyperref,xcolor}
\usepackage{amsmath,amssymb}
\usepackage{graphicx}
\usepackage{xcolor}
\usepackage{mathtools}
\usepackage{bm}
\usepackage{layouts}
\usepackage{pgf,tikz}
\usetikzlibrary{arrows}
\usetikzlibrary[patterns]
\definecolor{zzttqq}{rgb}{0.6,0.2,0}
\definecolor{cqcqcq}{rgb}{0.75,0.75,0.75}

\usepackage[margin=2.cm]{geometry}

\newcommand{\Z}{\ensuremath{\mathbb{Z}}}

\newcommand{\rd}{\ensuremath{\mathrm{d}}}

\providecommand{\abs}[1]{\lvert#1\rvert}

\newcommand{\expv}[1]{\left\langle#1\right\rangle}

\newcommand{\be}{\begin{equation}}
\newcommand{\ee}{\end{equation}}
\newcommand{\benn}{\nonumber\begin{equation}}
\newcommand{\eenn}{\nonumber\end{equation}}
\def\bea{\begin{eqnarray}} \def\eea{\end{eqnarray}}
\def\beann{\begin{eqnarray*}} \def\eeann{\end{eqnarray*}}
\def\lsim{\raise0.3ex\hbox{$<$\kern-0.75em\raise-1.1ex\hbox{$\sim$}}}
\def\gsim{\raise0.3ex\hbox{$>$\kern-0.75em\raise-1.1ex\hbox{$\sim$}}}

\DeclareMathOperator\Tr{Tr}
\renewcommand{\Re}{{\rm Re}}

\newcommand*\xbar[1]{%
  \hbox{%
    \vbox{%
      \hrule height 0.5pt 
      \kern0.3ex
      \hbox{%
        \kern-0.1em
        \ensuremath{#1}%
        \kern-0.1em
      }%
    }%
  }%
}

\graphicspath {{figures/}}

\begin{document}

\title{Mean distribution approach to spin and gauge theories}
\date{\today}
\author{Oscar Akerlund}
\affiliation{Institut f\"ur Theoretische Physik, ETH Z\"urich, CH-8093 Z\"urich, Switzerland}
\author{Philippe de Forcrand}
\affiliation{Institut f\"ur Theoretische Physik, ETH Z\"urich, CH-8093 Z\"urich, Switzerland}
\affiliation{CERN, Physics Department, TH Unit, CH-1211 Geneva 23, Switzerland}

\begin{abstract}
We formulate self-consistency equations for the distribution of links in spin models and of plaquettes
in gauge theories. This improves upon known mean-field, mean-link, and mean-plaquette approximations in such
that we self-consistently determine all moments of the considered variable instead of just the first.
We give examples in both Abelian and non-Abelian cases.

\end{abstract}

\maketitle

\section{Introduction}\noindent
\begin{tikzpicture}[overlay, remember picture]
\path (current page.north east) ++(-1,-1.5) node[below left] {\texttt{\footnotesize CERN-PH-TH/2015-291}};
\end{tikzpicture}
It is always of interest to think about methods that allow easy extraction of approximate results, even though the computer power
available for exact simulations is growing at an ever increasing pace. Mean-field methods are often qualitatively reliable
in their self-consistent determination of the long-distance physics, and have a wide range of applications, with spin models as
typical examples. For a gauge theory, formulated in terms of the gauge links, however, it is questionable what a \emph{mean link}
would mean, because of the local nature of the symmetry. This can be addressed by fixing the gauge, but the mean-field solution will
then in general depend on the gauge-fixing parameter. Nevertheless, Drouffe and Zuber developed techniques for a mean field treatment of general
Lattice Gauge Theories in~\cite{Drouffe:1983fv} and showed that for fixed $\beta d$, where $\beta$ is the inverse gauge coupling
and $d$ the dimension, the mean-field approximation can be considered
the first term in a $1/d$ expansion. They established that the mean field approximation can be thought of as a resummation of the
weak coupling expansion in a particular gauge and that there is a first order transition to a strong coupling phase at a critical value of $\beta$.
Since it becomes exact in the $d\to\infty$ limit, this mean field approximation can be used with some confidence in high-dimensional
models~\cite{Irges:2009bi}.

The crucial problem of gauge invariance was tackled and solved by Batrouni in a series
of papers~\cite{Batrouni:1982dx,Batrouni:1982bg}, where he first changed variables from gauge-variant links to gauge-invariant plaquettes.
The associated Jacobian is a product of lattice Bianchi identities,
which enforce that the product of the plaquette variables around an elementary cube is the identity element. In the Abelian case
this is easily understood, since each link occurs twice (in opposite directions) and cancels in this product, leaving the identity
element. In the non-Abelian case the plaquettes in each cube have to be parallel transported to a common reference point in order for
the cancellation to work. It is worth noting that in two dimensions there are no cubes so the Jacobian of the transformation is
trivial and the new degrees of freedom completely decouple (up to global constraints).

\begin{figure}[htp]
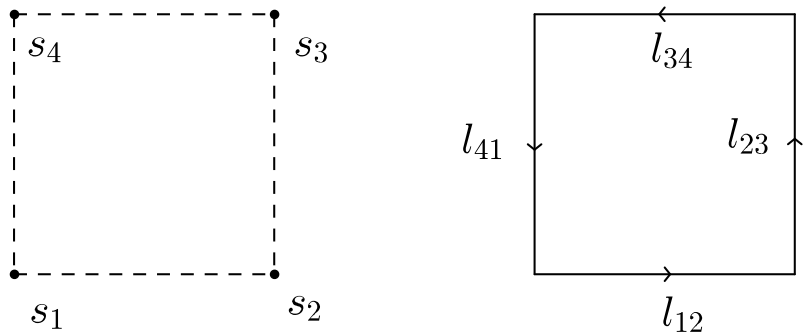

\centering
\includegraphics[width=0.6\linewidth]{{{../figures/bianchi_c}}}
\caption{The change of variables from spins $s_i$ (\emph{left panel})
  to links $l_{ij}$ (\emph{right panel}) that leads to the Bianchi identity
  $l_{12}l_{23}l_{34}l_{41} (=s_1s_2^\dagger s_2s_3^\dagger s_3s_4^\dagger s_4s_1^\dagger)=1$.}
  \label{fig:bianchi}
\end{figure}

This kind of change of variables can be performed for any gauge or spin model whose variables are elements of some group. Apart from gauge theories,
examples include $\Z_N$-spin models, $O(2)$- and $O(4)$-spin models and matrix-valued spin models.
In spin models, the change of variables is from spins to links and the Bianchi constraint dictates that the product of the links around an elementary
plaquette is the identity element. A visualization of the transformation and the Bianchi constraint for a $2d$ spin model is given in
Fig.~\ref{fig:bianchi}.

Let us review the change of variables for a gauge theory~\cite{Batrouni:1982bg}. The original variables are links. The new ones are plaquettes.
Under the action of the original symmetry of the model, the new variables transform within equivalence classes and it is possible
to employ a mean field analysis to determine the ``mean equivalence class''.
As usual we first choose a set of \emph{live} variables, which keep their original dynamics and interact with an external bath of
mean-valued fields. Interactions are generated through the Jacobian, which is a product of Bianchi identities represented by $\delta$-functions
\begin{equation}
  \label{eq:bianchi_id}
  \delta\left(\prod_{P\in\partial C}U_P-1\right),
\end{equation}
where $P$ denotes a plaquette and $\partial C$ denotes the oriented boundary of the elementary cube $C$. The $\delta$-functions can be
represented by a character expansion in which we can replace the characters at the external sites by their expectation, or mean, values.
Upon truncating the number of representations, this yields a closed set of equations in the expectation values which can be solved numerically.
The method can be systematically improved by increasing the number of representations used and the size of the live domain.

While this method works surprisingly well, even at low truncation, it determines the expectation value of the plaquette in only a few
representations. Here, we propose a method that self-consistently determines the complete distribution of the plaquettes (or links) and thus the
expectation value in all representations. This is due to an exact treatment of the lattice Bianchi identities which does not rely on
a character expansion. The only approximation then lies in the size of the live domain which can be systematically enlarged, as in any mean field
method. It is worth noting that our method works best for small $\beta$ and low dimensions: it does not become exact in the infinite dimension
limit. In this way it can be seen as complementary to the mean field approach of~\cite{Drouffe:1983fv}. We will also see that the mean distribution
approach proposed here actually works rather well for both small and large $\beta$.

The paper is organized as follows. In section~\ref{sec:method} we describe the method in general terms and compare it to the mean field,
mean link and mean plaquette methods before describing more detailed treatments of spin models and gauge theories in sections~\ref{sec:spin}~and~
\ref{sec:gauge} respectively. Finally, we draw conclusions in section~\ref{sec:conclusions}.

\section{Method}\label{sec:method}\noindent

\subsection{Mean Field Theory}\noindent
Let us for completeness give a very brief reminder of standard mean field theory. Consider for
definiteness a lattice model with a single type of variables $s$ which live on the lattice sites.
The lattice action is assumed to be translation invariant and of the form
\begin{equation}
  \label{eq:mf_action_orig}
  S = -\frac{1}{2}\sum_{i,j}J_{\abs{i-j}}s^\dagger_is_j + \sum_iV(s_i),
\end{equation}
where $i,j$ labels the lattice sites and $V(s)$ is some local potential. Let us now split the
original lattice into a live domain $D$ and an external bath $D^c$. The variables
$\{s_i\mid i\in D^c\}$ all take a constant ``mean'' value $\xbar{s}$. The mean field action
then becomes (up to a constant)
\begin{equation}
  \label{eq:mf_action_mf}
  S_\text{MF} = -\frac{1}{2}\sum_{i,j\in D}J_{\abs{i-j}}s^\dagger_is_j + \sum_{i\in D}\left(V(s_i)-\sum_{j\in D^c}J_{\abs{i-j}}s_i^\dagger\xbar{s}\right),
\end{equation}
where $\xbar{s}$ is determined by the self-consistency condition that the average value of $s$
in the domain $D$ is equal to the average value in the external bath,
\begin{equation}
  \label{eq:sc_mf}
  \expv{s} = \frac{\displaystyle\int\prod_{i\in D}\rd{}s_i\, s_i e^{-S_\text{MF}}}
  {\displaystyle\int\prod_{i\in D}\rd{}s_i\, e^{-S_\text{MF}}} \overset{!}{=} \xbar{s}.
\end{equation}
Once $\xbar{s}$ has been determined the mean field action~\eqref{eq:mf_action_mf} can be used
to measure other observables local to the domain $D$.

\subsection{Mean Distribution Theory}\noindent
To generalize the mean field approach we relax the condition that the fields at the live sites interact only with the mean
value of the external bath. Instead, the fields in the external bath are allowed to vary and take different values distributed
according to a \emph{mean distribution}. The self-consistency condition is thus that the distribution of the variables in the
live domain equals the distribution in the bath.

Consider a real scalar theory for illustration purposes. Starting from the action
\begin{equation}
  \label{eq:act_scalar}
  S = -2\kappa\sum_{\expv{i,j}}\phi_i\phi_j + \sum_i V\left(\phi_i\right),
\end{equation}
with nearest neighbor coupling $\kappa$ and a general on-site potential $V$, we expand the field $\phi\equiv\delta\phi+\xbar{\phi}$ around
its mean value $\xbar{\phi}$ and integrate out all the fields except the field at the origin $\phi_0=\xbar{\phi}+\delta\phi_0$ and its nearest neighbors,
denoted $\phi_i$, $i=1,\ldots,z$, where $z$ is the coordination number of the lattice. The partition function can then be written
\begin{equation}
  \label{eq:Z_scalar}
  Z = \int\rd\phi_0\,e^{-V\left(\phi_0\right)+2z\kappa\bar{\phi}\delta\phi_0}\int\prod_{i=1}^z\rd\delta\phi_i\,p_J(\delta\phi_1,\ldots,\delta\phi_z)
  e^{2\kappa\delta\phi_0\sum_{i=1}^z\delta\phi_i},
\end{equation}
where $p_J(\delta\phi_1,\ldots,\delta\phi_z)$ is a joint distribution function for the fields around the origin and absorbs everything not
explicitly depending of $\delta\phi_0$ into its normalization. So far everything
is exact and, given a way to compute $p_J$, we could obtain all local observables, for example $\expv{\phi_0^n}$. Now, $p_J$ is in general
not known, so we will have to make some ansatz and determine the best distribution compatible with this ansatz. In standard mean field
theory the ansatz is $p_J(\delta\phi_1,\ldots,\delta\phi_z)=\prod_{i=1}^z\delta(\delta\phi_i)$ and only $\xbar{\phi}$ is left to be
determined as explained above. In the mean distribution approach we will assume that the distribution is a \emph{product} distribution
$p_J(\delta\phi_1,\ldots,\delta\phi_z)=\prod_{i=1}^zp(\delta\phi_i)$ and determine $p$ self-consistently to be equal to the distribution
of $\delta\phi_0$, i.e.
\begin{equation}
  \label{eq:dist_scalar}
 p(\delta\phi_0) = \frac{1}{Z}e^{-V\left(\delta\phi_0+\bar{\phi}\right)+2z\kappa\bar{\phi}\delta\phi_0}\left(\expv{e^{2\kappa\delta\phi_0\delta\phi_i}}_{p(\delta\phi_i)}\right)^z,
\end{equation}
where $\expv{f(\phi)}_{p(\phi)}=\int\rd\phi\,p(\phi)f(\phi)$. The mean value $\xbar{\phi}$ has to be adjusted such that the distribution $p$
has zero mean. After $p$ and $\xbar{\phi}$ have been determined any observable, even observables extending outside the live domain, can be extracted
under the assumption that every plaquette is distributed according to $p$. Local observables are given by simple expectation values with respect
to the distribution $p$.

This strategy can also be applied to spin and gauge models, taking as variables the links and plaquettes respectively, as discussed
in the introduction. For a gauge theory, the starting point is the partition function in the plaquette formulation
\begin{equation}
  \label{eq:z_plaq}
  Z = \int\prod_P\rd U_P\,\prod_C\delta\left(\;\,\prod_{\mathclap{P\in\partial C}}U_P-1\right)e^{-S[U_P]},
\end{equation}
where $S[U_p]$ is any action which is a sum over the individual plaquettes, for example the Wilson action
$S[U_P] = \beta\sum_P(1-\mathrm{ReTr}U_P)$, or a topological action \cite{Bietenholz:2010xg,Akerlund:2015zha} where the action is
constant but the traces of the plaquette variables are limited to a compact region around the identity.

\begin{figure}[htp]
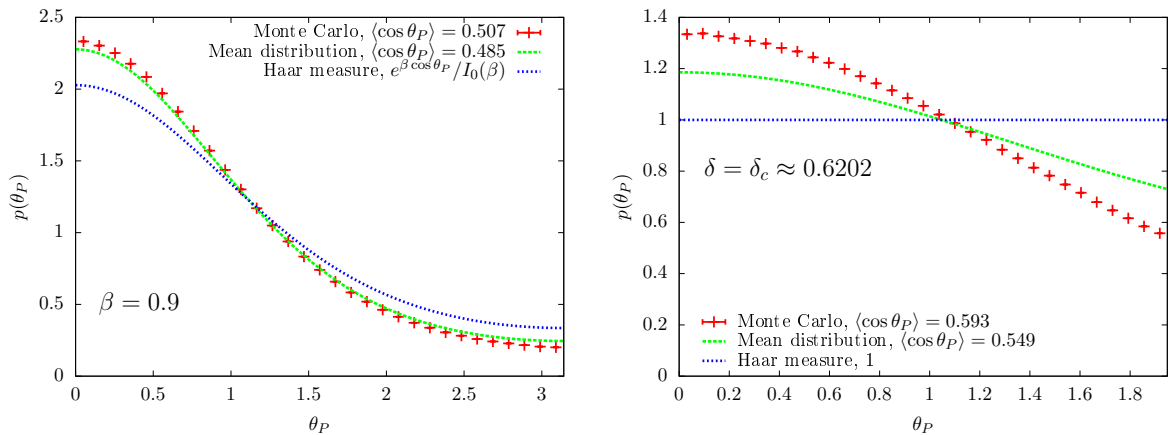

\centering
\includegraphics[width=0.45\linewidth]{{{../figures/4d_u1_beta_dist}}}
\includegraphics[width=0.45\linewidth]{{{../figures/4d_u1_rest_dist}}}
\caption{The distribution of plaquettes angles $p(\theta_P)$ in the $4d$ $U(1)$ lattice gauge theory with the Wilson action close to the critical
coupling (\emph{left panel}) and with the topological action at the critical restriction $\delta_c$ (\emph{right panel}) obtained by Monte Carlo
on an $8^4$ lattice and by the mean distribution approach, together with the Haar measure.}
  \label{fig:4d_u1_rest_dist}
\end{figure}

The difference to the mean plaquette method is that it is not assumed that the external plaquettes take some average value,
but rather that they are distributed according to a mean distribution. More specifically, we assume that there exists
a mean distribution for the real part of the trace of the plaquettes and that the other degrees of freedom are uniformly
distributed with respect to the Haar measure. Such a distribution must exist and it can be measured for example by Monte Carlo
simulations. For definiteness let us consider compact $U(1)$ gauge theory with a single plaquette $P_0$ as the live domain.
The plaquette variables $U_P = e^{i\theta_P}\in U(1)$ can be represented with a single real parameter $\theta_P\in[0,2\pi]$
and the real part of the trace is $\cos\theta_P$. Our goal is to obtain an approximation to the distribution
$p\left(\cos\theta_{P_0}\right)$, or equivalently $p\left(\theta_{P_0}\right) = Z\left(\theta_{P_0}\right)/Z$, where
\begin{align}
  Z(\theta_{P_0}) &= e^{-S[U_{P_0}]}\int\!\!\prod_{P\neq P_0}\!\!\rd U_P\,e^{-S[U_P]}\prod_C
  \delta\left(\;\;\prod_{\mathclap{P'\in\partial C}}U_P'-1\right),\label{eq:Z_th}\\
  Z &= \int\!\!\rd U_{P_0}\,Z(\theta_{P_0}).
\end{align}
To obtain a finite number of integrals we now make the approximation that all plaquettes which do not share a cube with $P_0$
are independently distributed according to some distribution $p(\theta)$. Clearly this neglects some correlations among the plaquettes but
this can be improved by taking a larger live domain. Again, let $C$ denote an elementary cube with boundary $\partial C$ and $P$ denote
a plaquette. We define
\begin{align}
U_C &\equiv \prod_{P\in\partial C}U_P,\\
\mathcal{C}_0 &\equiv \{C\mid P_0\in\partial C\},\\
\mathcal{P}_\mathcal{C} &\equiv \{P\mid \exists C \in \mathcal{C}_0:\,P\in\partial C,\,P\neq P_0\},
\end{align}
i.e. $\mathcal{C}_0$ is the set of all cubes containing $P_0$, and $\mathcal{P}_\mathcal{C}$ is the set of plaquettes, excluding $P_0$, making up $\mathcal{C}_0$.
The sought distribution is then determined by the self-consistency equation
\begin{equation}
  \label{eq:prob_th}
  p\left(\theta_{P_0}\right) = \frac{e^{-S\left[U_{P_0}\right]}\displaystyle\int\displaystyle\prod_{\mathclap{P\in \mathcal{P}_\mathcal{C}}}\rd U_P\,p\left(\theta_P\right)\displaystyle\prod_{\mathclap{C\in \mathcal{C}_0}}\delta\left(U_C-1\right)}
  {\displaystyle\int\!\!\rd U_{P_0}\,e^{-S\left[U_{P_0}\right]}\displaystyle\int\displaystyle\prod_{\mathclap{P\in \mathcal{P}_\mathcal{C}}}\rd U_P\,p\left(\theta_P\right)\displaystyle\prod_{\mathclap{C\in\mathcal{C}_0}}\delta\left(U_C-1\right)}.
\end{equation}
This self-consistency equation is solved by iterative substitution: given an initial guess for the distribution $p^{(0)}\left(\theta_{P_0}\right)$,
it is a straightforward task to integrate out the external
plaquettes and obtain the next iterate $p^{(1)}\left(\theta_{P_0}\right)$ from eq.~\eqref{eq:prob_th}, and to iterate the procedure until a fixed point is reached, i.e.
$p^{(n+1)}\left(\theta_{P_0}\right)=p^{(n)}\left(\theta_{P_0}\right)$. This is a functional equation, which is solved numerically by replacing the distribution
$p$ by a set of values on a fine grid in $\theta_P$ or by a truncated expansion in a functional basis. In this paper we have chosen to discretize the distribution
on a grid. As mentioned above, this can be done in a completely
analogous way also for spin models and for different types of actions. In Fig.~\ref{fig:4d_u1_rest_dist} we compare the distributions of plaquettes in the
$4d$ $U(1)$ lattice gauge theory with the Wilson action close to the critical coupling (\emph{left panel}) and with the topological action at
the critical restriction $\delta_c$ (\emph{right panel}), obtained by Monte Carlo on an $8^4$ lattice and by
the mean distribution approach with the normalized action $e^{\beta\cos\theta_P}$. Below we give more details for a selection of models along with numerical results.

\section{Spin models}\label{sec:spin}
We will start by applying the method to a few spin models, namely $\Z_2$, $\Z_4$ and the U(1) symmetric $XY$-model
and we will explain the procedure as we go along. Afterwards, only minor adjustments are needed in order to treat gauge theories.
We will derive the self-consistency equations in an unspecified number of dimensions although graphical illustrations will be given
in two dimensions for obvious reasons.

Let us start with an Abelian spin model with a global $\Z_N$ symmetry. The partition function is given by
\begin{equation}
  \label{eq:pf_zn}
  Z = \sum_{\{s\}}\exp\left(\beta\sum_{\expv{i,j}}\Re\, s_is_j^\dagger\right),
\end{equation}
where $s_i = e^{i\frac{2\pi}{N}n_i},\,n_i\in\{1,\cdots,N\} (\in\Z_N)$. In the usual mean field approach we would self-consistently determine
the mean value of $s_i$ by letting one or more live sites fluctuate in an external bath of mean valued spins.
However, Batrouni~\cite{Batrouni:1982dx,Batrouni:1985dp} noticed that by self-consistently determining
the mean value of the \emph{links}, or internal energy, $U_{ij}\equiv s_is_j^\dagger$, much better estimates of for example the critical temperature
could be obtained for a given live domain. Thus, we first change variables from spins to links. The Jacobian of this change of
variables is a product of lattice Bianchi identities, $\delta\left(U_P-1\right)$, one for each plaquette~\footnote{On a periodic lattice there
are also global Bianchi identities but they play no role here.}. This can be verified by introducing the link variables $U_{ij}$ via
$\int\rd U_{ij}\,\delta\left(U_{ij}s_js_i^\dagger-1\right)$ and integrating out the spins in a pedestrian manner. Since the Boltzmann weight factorizes
over the link variables, all link
interactions are induced by the Bianchi identities and hence the transformation trivially solves the one dimensional spin chain where there are
no plaquettes~\footnote{Up to a global constraint in the case of periodic boundary conditions}.

As mentioned above, each $\delta$-function can be represented by a sum over the characters of all the irreducible representations of the group.
For $\Z_N$ this is merely a geometric series, $\delta\left(U_P-1\right) = \frac{1}{N}\sum_{n=0}^{N-1}U_P^n$. Since only the real part enters in the
action it is convenient to reshuffle the sum so that we sum only over real combinations of the variables,
\begin{equation}
  \label{eq:delt_ZN}
  \delta\left(U_P-1\right) \propto 1+U_P^{N/2}\delta^N_\text{even}+\sum_{n=1}^{\mathclap{\left\lfloor\frac{N-1}{2}\right\rfloor}}\left(U_P^n+U_P^{-n}\right),
\end{equation}
where $\delta^N_\text{even}$ is $1$ if $N$ is even and $0$ otherwise.

\begin{figure}[htp]
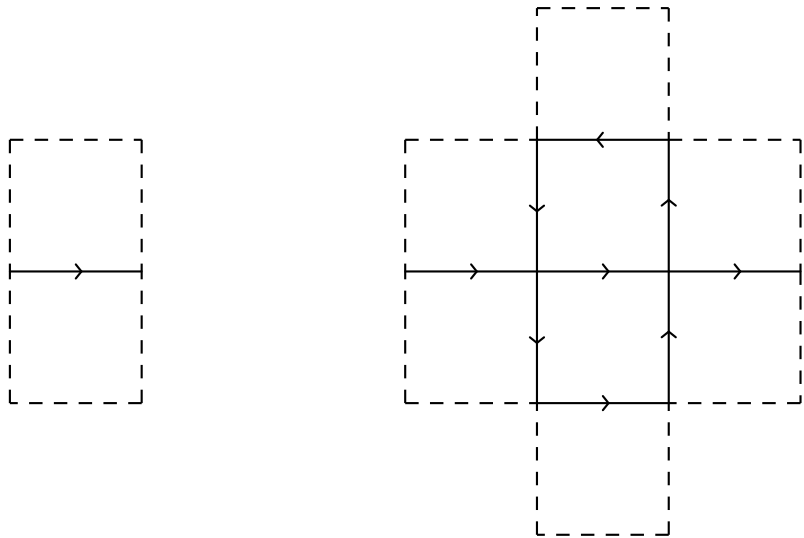

\centering
\includegraphics[width=0.6\linewidth]{{{../figures/live_domains_c}}}
\caption{Two choices of domains of live links for $2d$ spin models.
  The live links are denoted by the solid lines, whereas the dashed lines denote
  links which are assumed to take mean values or to be distributed according to
  the mean distribution. The left panel shows the unique domain with one live
  link and the right panel shows one of many domains with nine live links.}
  \label{fig:live_domains}
\end{figure}

The next step is to choose a domain of live links. In this step, imagination is the limiting factor; for a given number of live
links there can be many different choices and it is not known to us if there is a way to decide which is the optimal one.
The simplest choice is of course to keep only one link alive but in our $2d$ examples we will make use also of
a nine-link domain~\cite{Batrouni:1985dp} to see how the results improve with larger domains. These two domains are shown
in the left (one link) and right (nine links) panels of Fig.~\ref{fig:live_domains}. In the case of a single live link, there
are $2(d-1)$ plaquettes and thus there are $2(d-1)$ $\delta$-functions of the type in eq.~\eqref{eq:delt_ZN}.

\subsection{Mean link approach}\noindent
Let us for simplicity consider the case of one live link, denoted $U_0$. The external links, denoted $U_k$ by some enumeration $ij\to k$, are fixed to
the mean value by demanding that $U_k^n=U_k^{-n}=\expv{U}^n,\, \forall k\neq0$.
Each plaquette containing the live link also contains three external links, and the $\delta$-function eq.~\eqref{eq:delt_ZN} becomes
\begin{equation}
  \delta\left(U_P-1\right) \propto 1+\expv{U}^{3N/2}(-1)^{n_0}\delta^N_\text{even}
  +2\sum_{n=1}^{\mathclap{\left\lfloor\frac{N-1}{2}\right\rfloor}}\expv{U}^{3n}\cos\frac{2\pi n_0 n}{N}.  \label{eq:delt_ZN_av}
\end{equation}
For large $N$ it is best to perform the sum analytically to obtain (for $N=2M$)
\begin{equation}
  \label{eq:delt_ZN_rs}
  \delta\left(U_P-1\right) \propto \frac{1-(-1)^{n_0}\expv{U}^{3M}}{1+\expv{U}^6-2\expv{U}^3\cos\frac{\pi n_0}{M}}.
\end{equation}
For U(1) we define $\frac{\pi n_0}{M} = \theta_0$ as $M\to\infty$ and since $\expv{U}<1$ we get
\begin{equation}
  \label{eq:delt_ZN_u1}
  \delta\left(U_P-1\right) \propto \left(1+\expv{U}^6-2\expv{U}^3\cos\theta_0\right)^{-1},
\end{equation}
which can efficiently be dealt with by numerical integration. The partition functions for the single
live link for $\Z_2$, $\Z_4$ and $U(1)$~\footnote{The $U(1)$ Wilson action is defined by $\delta=\pi,\beta\neq0$ and the
  topological action by $\delta<\pi,\beta=0$.} spin models then become
\begin{align}
  Z_{\Z_2}&\propto\sum_{U_0=\pm1}e^{\beta U_0}\left(1+\expv{U}^3U_0\right)^{2(d-1)},\\
  Z_{\Z_4}&\propto\sum_{n_0=0}^3e^{\beta\cos\frac{\pi n_0}{2}}\left(1+\expv{U}^6(-1)^{n_0}+2\expv{U}^3\cos\frac{\pi n_0}{2}\right)^{2(d-1)},\\
  Z_{U(1)}&\propto\displaystyle\int\limits_{\mathclap{-\delta}}^\delta \rd\theta\,e^{\beta \cos\theta}\left(1+\expv{U}^6-2\expv{U}^3\cos\theta\right)^{-2(d-1)}.\label{eq:u1_spin}
\end{align}
In the $U(1)$ case, eq.~\eqref{eq:u1_spin} applies both to the standard action $(\beta\geq0,\delta=\pi)$ and to the topological action
$(\beta=0,\delta\leq\pi)$.

\subsection{Mean distribution approach}\noindent
In the mean distribution approach we sum over the external links assuming they each obey a mean distribution $p(U)$,
for which a one-to-one mapping to the set of moments $\{\expv{U^n}\}$ exists. The difference between the two methods
becomes apparent when expressed in terms of the moments, which are obtained by integrating the distributions of the external links
against the $\delta$-function given by the Bianchi constraint in eq.~\eqref{eq:delt_ZN}
\begin{equation}
  \label{eq:dela_dist}
  \sum_{\mathclap{\{U_1,U_2,U_3\}}}p(U_1)p(U_2)p(U_3)\delta(U_P-1)=
  1+\expv{U^{N/2}}^3U_0^{N/2}\delta^N_\text{even} + 2\sum_{n=1}^{\mathclap{\left\lfloor\frac{N-1}{2}\right\rfloor}}\expv{U^n}^3\cos\frac{2\pi n_0n}{N}.
\end{equation}
Comparing to eq~\eqref{eq:delt_ZN_av}, we see that for $N\leq3$ there is only one moment and the two methods are thus equivalent,
but for larger $N$ the mean link approach
makes the approximation $\expv{U^n}=\expv{U}^n$ whereas the mean distribution approach treats all moments correctly.

Thus, for small $N$ we do not expect much difference between the two approaches, and this is indeed confirmed by explicit calculations.
For $U(1)$, however, there are infinitely many moments which are treated incorrectly by the mean link approach and this renders the mean
distribution approach conceptually more appealing.

By using the Bianchi identities, one link per plaquette can be integrated out, giving
\begin{equation}
  Z_{U(1)}=\int\limits_{\mathclap{-\delta}}^\delta \rd\theta\,e^{\beta \cos\theta}
  \left(\int\limits_{-\delta}^{\delta}\rd\theta_1\rd\theta_2\,p(\theta_1)p(\theta_2)\sum_{n=-2}^2p(2\pi{}n-\theta-\theta_1-\theta_2)\right)^{\mathrlap{2(d-1)}}.
\end{equation}
It is often convenient not to work solely with distributions of single links, but also of multiple links, which are
defined in the obvious way,
\begin{equation}
  \label{eq:dist_multi}
  p_N(\Theta)\equiv\int\prod_{i=1}^N\rd\theta_i\,p(\theta_i)\delta\left(\sum_{i=1}^N\theta_i-\Theta\right),
\end{equation}
and can efficiently be calculated recursively. The above partition function then simplifies slightly to
\begin{equation}
  Z_{\text{U(1)}}=\int\limits_{\mathclap{-\delta}}^\delta \rd\theta\,e^{\beta \cos\theta}\left(\int\limits_{-2\delta}^{2\delta}\rd\Theta\,p_2(\Theta)
    \sum_{n=-2}^2p(2\pi{}n-\theta-\Theta)\right)^{2(d-1)}.
\end{equation}

\begin{figure}[htp]
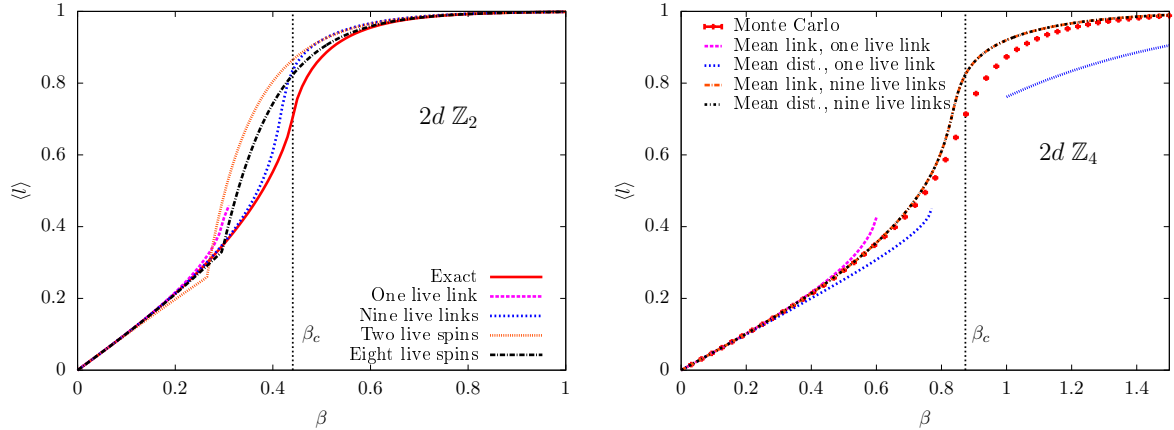

\centering
\includegraphics[width=0.45\linewidth]{{{../figures/2d_ising}}}
\includegraphics[width=0.45\linewidth]{{{../figures/2d_z4}}}
\caption{(\emph{left}) Mean-field and mean-link approximation in the $2d$ Ising model for two choices of live domains.
  (\emph{Right}) Mean-link and mean-distribution in the  $2d$ $\Z_4$ model. In the Ising case, mean-link and mean-distribution are equivalent.}
  \label{fig:2d_ising_z4}
\end{figure}

\begin{figure}[htp]
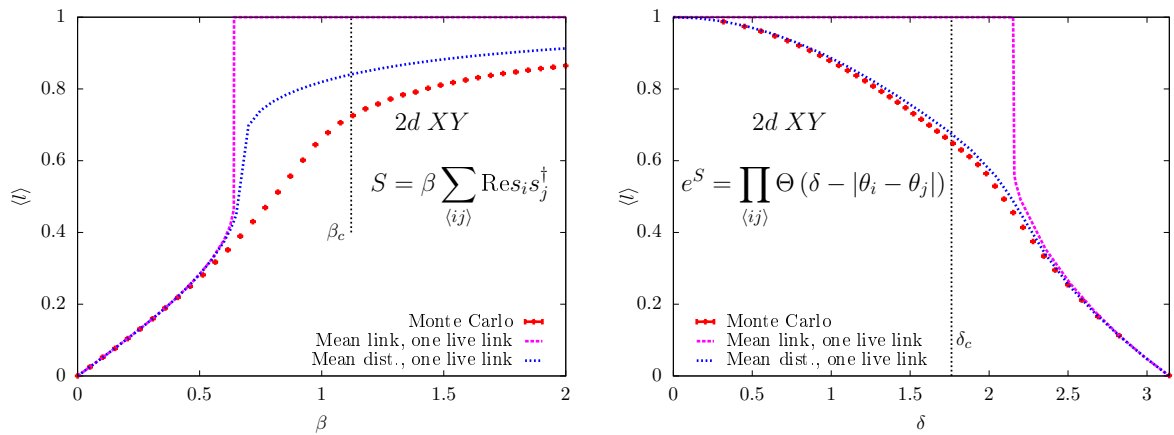

\centering
\includegraphics[width=0.45\linewidth]{{{../figures/2d_xy_beta}}}
\includegraphics[width=0.45\linewidth]{{{../figures/2d_xy_rest}}}
\caption{The mean link in the $2d$ $XY$ spin model as a function of
  the Wilson coupling $\beta$ (\emph{left panel}) and of the restriction $\delta$
  (\emph{right panel}) from Monte Carlo, from the mean link and from
  the mean distribution methods.}
  \label{fig:2d_u1}
\end{figure}

In Figs.~\ref{fig:2d_ising_z4}~and~\ref{fig:2d_u1} we show results for $2d$ $\Z_2$, $\Z_4$ and $U(1)$ spin models, the latter for the Wilson action
$S=\beta\sum_{\expv{ij}}\Re\, s_is_j^\dagger$ and the topological action $e^S=\prod_{\expv{ij}}\Theta\left(\delta-\abs{\theta_i-\theta_j}\right)$.
Note the remarkable accuracy of the mean distribution approach in the latter case, even when there is only one live link.

\section{Gauge theories}\label{sec:gauge}
\begin{figure}[htp]
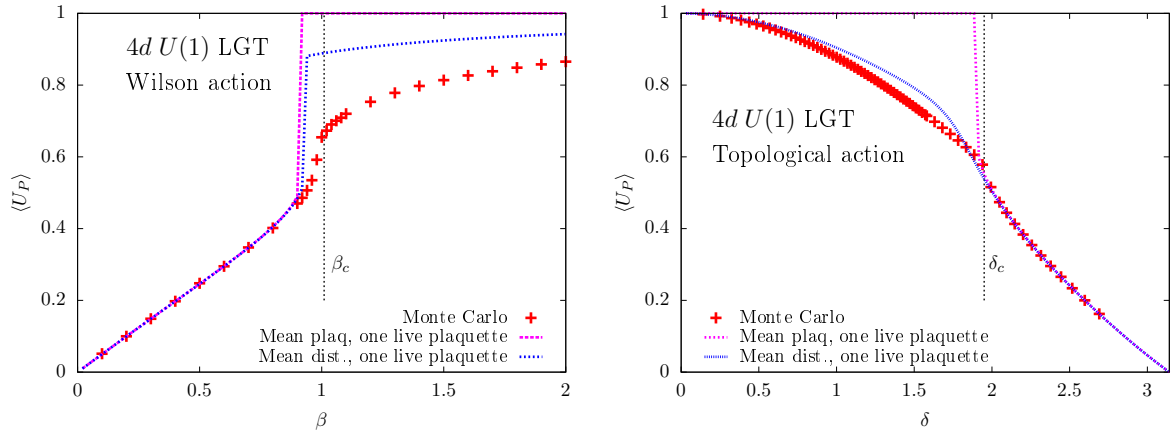

\centering
\includegraphics[width=0.45\linewidth]{{{../figures/4d_u1_beta}}}
\includegraphics[width=0.45\linewidth]{{{../figures/4d_u1_rest}}}
\caption{The mean plaquette in the $4d$ $U(1)$ gauge theory as a function of
  the Wilson coupling $\beta$ (\emph{left panel}) and the restriction $\delta$
  (\emph{right panel}) from Monte Carlo, and from the mean plaquette and
  the mean distribution methods.}
  \label{fig:4d_u1}
\end{figure}

To extend the formalism from spin models to gauge theories, we merely have to change from links and plaquettes to plaquettes and cubes.
The partition function for a $U(1)$ gauge theory analogous to eq.\eqref{eq:u1_spin} becomes
\begin{equation}\label{eq:u1_gauge}
  Z_{\text{U(1)}}=\int\limits_{\mathclap{-\delta}}^\delta \rd\theta\,e^{\beta \cos\theta}\left(1+\expv{U}^{10}-2\expv{U}^5\cos\theta\right)^{-2(d-2)}
\end{equation}
in the mean plaquette approach and
\begin{equation}
  Z_{\text{U(1)}}=\int\limits_{\mathclap{-\delta}}^\delta \rd\theta\,e^{\beta \cos\theta}\left(\int\limits_{-4\delta}^{4\delta}\rd\Theta\,p_4(\Theta)\sum_{n=-3}^3p(2\pi{}n-\theta-\Theta)\right)^{2(d-2)}
\end{equation}
in the mean distribution approach. Results for $d=4$ are shown in Fig.~\ref{fig:4d_u1}
for the Wilson action (\emph{left panel}) and for the topological action (\emph{right panel}).

\begin{figure}[htp]
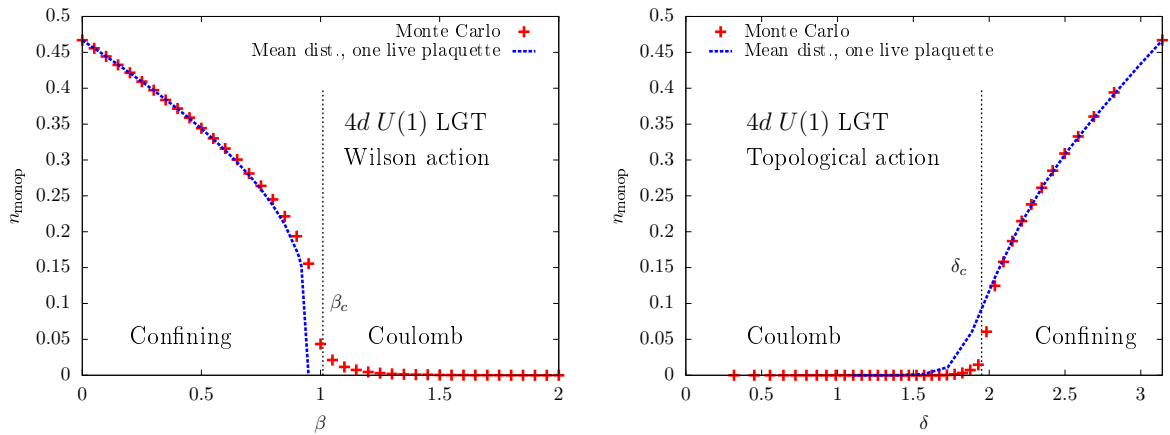

\centering
\includegraphics[width=0.45\linewidth]{{{../figures/4d_u1_beta_monop}}}
\includegraphics[width=0.45\linewidth]{{{../figures/4d_u1_rest_monop}}}
\caption{The monopole density in the $4d$ $U(1)$ gauge theory as a function of
  the Wilson coupling $\beta$ (\emph{left panel}) and the restriction $\delta$
  (\emph{right panel}) from Monte Carlo and the mean distribution method.}
  \label{fig:4d_u1_monop}
\end{figure}

Another nice feature of the mean distribution approach is that other observables become available, like for instance the monopole density
in the $U(1)$ gauge theory, under the assumption that each plaquette is distributed according to the mean distribution $p$.
A cube is said to contain $q$ monopoles if the sum of its outward
oriented plaquette angles sums up to $2\pi{}q$. Given the distribution $p(\theta)$ of plaquette angles the
(unnormalized) probability $p_q$ of finding $q$ monopoles in a cube is given by
\begin{equation}
  \label{eq:monopole_prob}
  p_q = \int\prod_{i=1}^6\rd \theta_i\, p(\theta_i) \delta\left(\sum_{i=1}^6\theta_i - 2q\pi\right),\, q\in\{-2,-1,0,1,2\}
\end{equation}
and the monopole density $n_{\rm monop}$ is given by
\begin{equation}
  \label{eq:nmonop}
  n_{\rm monop} = \frac{2p_1 + 4p_2}{p_0+2p_1+2p_2}.
\end{equation}
In Fig.~\ref{fig:4d_u1_monop} we show the monopole densities for $4d$ $U(1)$ gauge theory as obtained by
Monte Carlo simulations and by the mean distribution approach. Note that the monopole extends outside of the domain
of a single live plaquette, which was used to determine the mean distribution $p$. The left panel shows results for the Wilson
action and in the right panel the topological action is used.

We can also treat $SU(2)$ Yang-Mills theory without much difficulty. For the mean plaquette approach we need the character expansion of
the $\delta$-function
\begin{equation}
  \label{eq:delt_su2}
  \delta\left(U_C-1\right) \propto \sum_{n=0}^\infty(n+1)\frac{\sin(n+1)\theta_C}{\sin\theta_C},
\end{equation}
where $\theta_C$ is related to the trace of the cube matrix $U_C$ through $\text{Tr}U_C = 2\cos\theta_C$.

In the mean plaquette approach we again make the substitution $U_C\to U_0\expv{U}^5$ in the case of a single live plaquette. The above delta
function then becomes
\begin{align}
  \label{eq:delt_su2_mp}
  \delta\left(U_0\expv{U}^5-1\right) &\propto \sum_{n=0}^\infty\expv{U}^{5n}(n+1)\frac{\sin(n+1)\theta_0}{\sin\theta_0}\nonumber\\
  &\propto  \left(1+\expv{U}^{10}-2\cos\theta_0\expv{U}^5\right)^{-2}.
\end{align}
For $SU(2)$, the analogue of a restriction $\delta$ on the plaquette angle is a restriction on the trace of the plaquette matrix to the domain $[2\alpha,2]$, where 
$-1\leq\alpha<1$. If we define $a_0 \equiv \frac{1}{2}\Tr U_0=\cos\theta_0$ the approximate $SU(2)$ partition function can be
written~\footnote{The $SU(2)$ Wilson action is defined by $\alpha=-1,\beta\neq0$ and the topological action by $\alpha>-1,\beta=0$.}
in a way very similar to the $U(1)$ partition function~\eqref{eq:u1_gauge}
\begin{equation}
  Z_{\text{SU(2)}}=\int\limits_{\alpha}^1 \rd a_0\,\sqrt{1-a_0^2}\,e^{\beta a_0}\left(1+\expv{U}^{10}-2\expv{U}^5a_0\right)^{-4(d-2)},
\end{equation}
from which $\expv{U}$ can be easily obtained as a function of $\alpha$ and $\beta$.

The mean distribution approach works in a completely analogous way as for $U(1)$, but let us go through the details anyway, since there
are now extra angular variables to be integrated out. The starting point is again an elementary cube on the lattice. Five of the cubes faces
have their trace distributed according to the distribution $p(a_0)$ and we want to calculate the distribution of the sixth face compatible with
the Bianchi identity $U_C=1$. In other words, taking $U_6$ as the live plaquette, we want to evaluate
\begin{equation}
  \tilde{p}(a_{0,6})\propto \left.\int\rd \Omega_6 \int\prod_{i=1}^5\left\{\rd U_i\frac{p(a_{0,i})}{\sqrt{1-a_{0,i}^2}}\right\}
    \delta\left(\prod_{i=1}^6U_i-1\right)\right\rvert_{\text{Tr}U_6=2a_{0,6}},
\end{equation}
where we have decomposed $U_6=\Omega_6 \hat{U}_6\Omega_6^\dagger$ with $\hat{U}_6$ a diagonal $SU(2)$ matrix with trace $2a_{0,6}$, i.e.
$\Omega_6$ is the angular part of $U_6$. The choice to include the measure factor $\sqrt{1-a_0^2}$ in the distribution is arbitrary but
convenient. To facilitate the calculation we recursively combine the product of four of the plaquette matrices into one matrix, $U_1U_2U_3U_4\to\tilde U$,
by pairwise convolution of distributions (with $p_1(a_0)\equiv{}p(a_0)$)
\begin{align}
  p_{2i}(\tilde{a}_0) &\propto \left.\int\rd\tilde{\Omega}\rd U_1 \rd U_2 \frac{p_i(a_{0,1})}{\sqrt{1-a_{0,1}^2}}\frac{p_i(a_{0,2})}{\sqrt{1-a_{0,2}^2}}
    \delta\left(U_1U_2\tilde{U}^\dagger-1\right)\right\rvert_{\text{Tr}\tilde{U}=2\tilde{a}_0}\nonumber\\
  &\propto \int\limits_{\alpha_i}^1\rd a_{0,1}\rd a_{0,2}\,p_i(a_{0,1})p_i(a_{0,2})\int\limits_{-1}^1\rd\cos\theta_{12}\,
  \delta\left(\tilde{a}_0-a_{0,1}a_{0,2}-\sqrt{1-a_{0,1}^2}\sqrt{1-a_{0,2}^2}\cos\theta_{12}\right)\\
  &=\int\limits_{\alpha_i}^1\rd a_{0,1}\rd a_{0,2}\frac{p_i(a_{0,1})p_i(a_{0,2})}{\sqrt{1-a_{0,1}^2}\sqrt{1-a_{0,2}^2}}
  \chi_{\abs{\tilde{a}_0-a_{0,1}a_{0,2}}\leq\sqrt{1-a_{0,1}^2}\sqrt{1-a_{0,2}^2}},\nonumber
\end{align}
where $\alpha_1 \equiv \alpha,\,\alpha_{2i}=\max(2\alpha_i-1,-1)$ and $\chi_A$ is the characteristic function on the domain $A$. The domain of integration in the
$(a_{0,1},a_{0,2})$-plane is simply connected with parametrizable boundaries and comes from the condition that the argument of the delta function has a zero
for some $\cos\theta_{12}\in[-1,1]$. We then obtain for the sought distribution
\begin{equation}
  \tilde{p}(a_{0,6})\propto \left.\int\rd \Omega_6 \int\rd U_5\frac{p(a_{0,5})}{\sqrt{1-a_{0,5}^2}}\int\rd \tilde{U}\frac{p_4(\tilde{a}_0)}{\sqrt{1-\tilde{a}_0^2}}\delta\left(\tilde{U}U_5U_6-1\right)\right\rvert_{\text{Tr}U_6=2a_{0,6}},
\end{equation}
where it is now easy to integrate out $\tilde{U}=U_6^\dagger U_5^\dagger$. If we denote by $\theta_{56}$ the angle between $U_5$ and $U_6$, the angular
integral over $\Omega_6$ contributes just a multiplicative constant and we obtain
\begin{equation}
  \tilde{p}(a_{0,6})\propto \int\rd a_{0,5}\rd\cos\theta_{56}\,p(a_{0,5})\frac{p_4\left(a_{0,5}a_{0,6}-\sqrt{1-a_{0,5}^2}\sqrt{1-a_{0,6}^2}\cos\theta_{56}\right)}
  {\sqrt{a_{0,5}a_{0,6}-\sqrt{1-a_{0,5}^2}\sqrt{1-a_{0,6}^2}\cos\theta_{56}}},
\end{equation}
which can be evaluated numerically in a straightforward manner. In the end, since there are $2(d-2)$ cubes sharing the plaquette $P_0$,
and since the a priori probability for $P_0$ to have trace $2a_0$ is $\sqrt{1-a_0^2}e^{\beta a_0}$, with respect to the uniform measure,
we obtain for one live plaquette
\begin{align}
  Z_{SU(2)}&=\int\limits_{\alpha}^1\rd a_{0}\,p(a_0)=\int\limits_{\alpha}^1\rd a_{0}\sqrt{1-a_0^2}e^{\beta a_0}\tilde{p}(a_0)^{2(d-2)}\nonumber\\
  &= \int\limits_{\alpha}^1\rd a_{0}\sqrt{1-a_0^2}e^{\beta a_0}\left(\int\limits_{\alpha}^1\rd x\, p(x)\rd\cos\theta\frac{p_4\left(a_0x-\sqrt{1-a_0^2}\sqrt{1-x^2}
        \cos\theta\right)}{\sqrt{a_0x-\sqrt{1-a_0^2}\sqrt{1-x^2}\cos\theta}}\right)^{2(d-2)},
\end{align}
which also defines the functional self-consistency equation for $p(a_0)$.

\begin{figure}[htp]
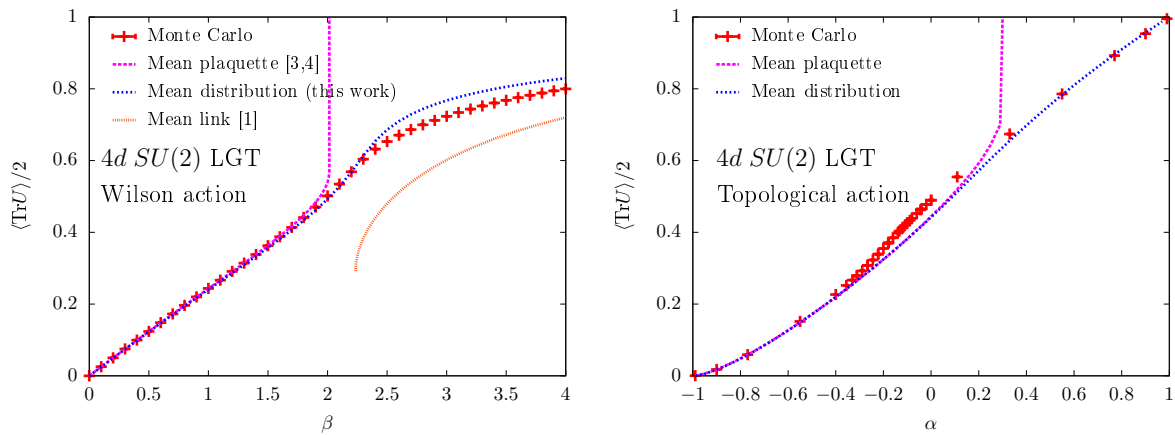

\centering
\includegraphics[width=0.45\linewidth]{{{../figures/su2_plaq_beta}}}
\includegraphics[width=0.45\linewidth]{{{../figures/su2_plaq_rest}}}
\caption{The average plaquette for the $SU(2)$ gauge theory as a function of the
  Wilson coupling $\beta$ (\emph{left panel}) and the restriction $\alpha$ (\emph{right panel})
  from Monte Carlo simulation, the mean plaquette method and the mean distribution method.
  For comparison the mean link result obtained with the formalism in~\cite{Drouffe:1983fv} is also shown in the left panel.}
\label{fig:su2}
\end{figure}

Results for the Wilson and topological actions can be seen in Fig.~\ref{fig:su2} in the left
and right panels, respectively~\footnote{Our results for the mean plaquette approach differ a little from those of~\cite{Batrouni:1982dx},
  because we imposed the Bianchi constraint exactly rather than truncating its character expansion.
  Surprisingly, truncation gives better results.}.

For $SU(3)$ one can proceed in an analogous manner, only the angular integrals are now more involved and the trace of the plaquette
depends on two diagonal generators so the resulting distribution function needs to be two dimensional.

\section{Conclusions}\label{sec:conclusions}\noindent
It has been shown before~\cite{Batrouni:1985dp} that determining a self-consistent mean-link gives a much better
approximation than the traditional mean-field. Furthermore, the symmetry-invariant mean link can be
generalized to a mean plaquette in gauge theories~\cite{Batrouni:1982dx}. Here, we have shown that the
approximation can be further improved by determining the self-consistent mean distribution of links
or plaquettes.
The extension from a self-consistent determination of the symmetry invariant
mean link or plaquette to a self-consistent determination of the entire
distribution of  links and plaquettes  is shown to improve upon the results
obtained by Batrouni in his seminal work~\cite{Batrouni:1982dx,Batrouni:1982bg}. Especially appealing is the fact that
the mean distribution approach yields a non-trivial result for the whole range
of couplings and not just in the strong coupling regime, which is sometimes the case
for the mean link/plaquette approach, or just in the weak coupling regime which is
accessible to the mean field treatment of~\cite{Drouffe:1983fv}. Indeed, the mean distribution approach
gives a nearly correct answer when the correlation length is not too large,
and by enlarging the live domain the exact result is approached systematically for any value of the coupling.
As the domain of live variables is enlarged, the mean link/plaquette and the mean distribution results
tend to approach each other but since determining the full mean distribution does not require much additional
computer time it should always be desirable to do so.

Furthermore, another appealing feature of the mean distribution approach is that once the
distribution has been self-consistently determined, other local observables,
like the vortex or monopole densities become readily available. Finally, the whole approach
applies to non-Abelian models as well.

\bibliography{\jobname.bib}

\end{document}